\documentclass[twocolumn,showpacs,preprintnumbers,amsmath,amssymb]{revtex4}

\usepackage{graphicx}
\usepackage{dcolumn}
\usepackage{bm}

\begin{document}


\title{Bandstructure Effects in Multiwall Carbon Nanotubes} 

\author{Bernhard Stojetz and Christoph Strunk}
\affiliation{Institute of Experimental and Applied Physics, University of Regensburg, 93040 Regensburg, Germany}
\author{Csilla Miko and Laszlo Forr\'{o}}
\affiliation{Institute of Physics of Complex Matter, FBS 
Swiss Federal Institute of Technology (EPFL), CH-1015 Lausanne, Switzerland}

\date{\today}

\begin{abstract}
We report conductance measurements on multiwall carbon nanotubes in a perpendicular magnetic field. A gate electrode with large capacitance is used to considerably vary the nanotube Fermi level. 
This enables us to search for signatures of the unique electronic band structure of the nanotubes in the regime of diffusive quantum transport. We find an unusual quenching of the magnetoconductance and the zero bias anomaly in the differential conductance at certain gate voltages, which can be linked to the onset of quasi-one-dimensional subbands.
\end{abstract}
\maketitle
Quantum transport in multiwall carbon nanotubes has been intensely studied in recent years \cite{DekkerReview,Schoenenberger}. Despite some indications of ballistic transport even at room temperature \cite{Frank,Urbina}, the majority of experiments revealed typical signatures of diffusive quantum transport in a magnetic field $B$ such as weak localization (WL), universal conductance fluctuations (UCF) and the $h/2e$-periodic Altshuler-Aronov-Spivak (AAS) oscillations \cite{Langer,Schoenenberger,LiuAvouris,Bachtold_AB}. These phenomena are caused by the Aharonov-Bohm phase, either by coherent backscattering of pairs of time-reversed diffusion paths (WL and AAS) or by interference of different paths (UCF). In addition, zero bias anomalies caused by electron-electron interactions in the differential conductance have been observed \cite{Bachtold_Tunnel}. In those experiments, the multiwall tubes seemed to behave as ordinary metallic quantum wires. On the other hand, bandstructure calculations for singlewall nanotubes predict strictly one-dimensional transport channels, which give rise to van Hove singularities in the density of states\cite{SaitoDresselhaus}. Experimental evidence for this has been obtained mainly by electron tunneling spectroscopy on single wall nanotubes \cite{TansDekker393}. In this picture of strictly one-dimensional transport a quasiclassical trajectory cannot enclose magnetic flux and no low-field magnetoconductance is expected. Hence, the question arises how the specific band structure is reflected in the conductance as well as in its quantum corrections and how those on first glance contradictory approaches can be merged into a consistent picture of electronic transport.\\ 
In this experiment, we use a strongly coupled gate, which is efficient enough to shift the Fermi level through several quasi-onedimensional subbands. At certain gate voltages, which can be associated with the bottoms of the subbands, we observe a strong suppression of both the magnetoconductance and the differential conductance.\\
\begin{figure}
\includegraphics{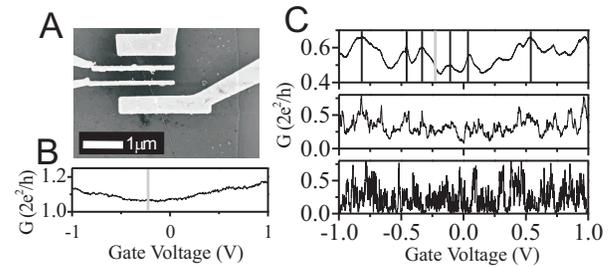}
\caption{\label{fig1}(A) SEM image of a typical sample: individual multiwall nanotubes are deposited on a prestructured Al gate electrode and contacted by four Au fingers, which are deposited on top of the tube. The electrode spacing is 300 nm. For the measurements, only the two inner electrodes are used. (B) Room temperature conductance of sample A as a function of gate voltage in units of the conductance quantum $2e^2/h$. The estimated position of the charge neutrality point corresponds to the minimum of conductance and is indicated by a grey line. (C) Same as in Fig. 1B, but for 10 K, 1 K and 30 mK (top to bottom). For the 10K curve, both the positions of the  charge neutrality point (grey line) and the regions of quenched magnetoconductance (black lines) as observed in Fig. 2 are indicated.}
\end{figure}
The samples were produced on top of thermally oxidized Silicon
wafers. First, Aluminium strips of 10 $\mu$m width and 40 nm thickness
were evaporated. Exposure to air provides an insulating native oxide
layer of a few nm thickness. These strips serve as a backgate for the
individual nanotubes, which are deposited from a chloroform suspension
in the next step. Electric contacts are defined by electron beam
lithography. After application of an oxygen plasma, 80 nm of Gold are
deposited. In this way we achieve typical resistances between \mbox{10
  k$\Omega$} and \mbox{30 k$\Omega$} at 4.2 K. The samples were
operated by a low frequency ac bias voltage and application of a dc
gate voltage $U_{\rm Gate}$ to the Aluminium layer. Up to gate
voltages of 3 V no leakage current between the gate and the tube was
observed ($I_{\rm Leak}<100$ fA). Typical breakdown voltages of the
gate oxide were 3-4 V. Two-terminal resistance measurements were
carried out for two samples, A and B. The lengths of the samples are
\mbox{5 $\mu$m} and \mbox{2 $\mu$m} and their diameters are \mbox{19
  nm} and \mbox{14 nm}, respectively. A scanning electron micrograph
of a typical sample is presented in Fig. 1A.\\
In order to characterize the dependence of the conductance of sample A on $U_{\rm Gate}$, a small ac bias voltage of 2 $\mu$V $\ll k_{\rm B}T$ was applied and the current was measured at  several temperatures $T$ (Fig. 1B,C). Fig. 1B shows the conductance $G$ as a function of gate voltage at 300 K. The corresponding curves for 10 K, 1 K and 30 mK are presented in Fig. 1C. The conductance at room temperature exhibits a shallow minimum located at \mbox{$U_{\rm Gate}\approx -$0.2 V}. When the Fermi level is tuned away from the charge neutrality point, more and more subbands can contribute to the transport and an increase of the conductance is expected. Thus we attribute the position of the conductance minimum to the charge neutrality point, where bands with positive energy are unoccupied while those with negative energies are completely filled \cite{Krueger}. This reveals the high efficiency of the gate as well as an intrinsic {\it n}-doping of the tube. The location of the minimum varied from sample to sample. We observed {\it p}- as well as {\it n}-doping at\mbox{ $U_{\rm Gate}=0$ V} in several samples. The $G(U_{\rm Gate})$ curves in Fig. 1C show an increasing amplitude of the conductance fluctuations as the temperature is lowered, while the average conductance decreases. This can be interpreted as a gradual transition from a coexistence of band structure effects, UCFs and charging effects at 10 K and 1 K to the dominance of Coulomb blockade at 30 mK. In contrast to experiments on clean single wall nanotubes, no periodic Coulomb oscillations are found. Instead, irregular peaks in conductance occur. It is likely that disorder induces a nonuniform series of strongly coupled quantum dots and that transport is governed by higher order tunneling processes \cite{Bockrath}.\\
Next, conductance traces $G(U_{\rm Gate})$ were recorded at several
 temperatures and in magnetic fields perpendicular to the tube
 axis. The result at a temperature of 10 K is displayed as a color
 plot in Fig. 2A.
\begin{figure*}
\includegraphics{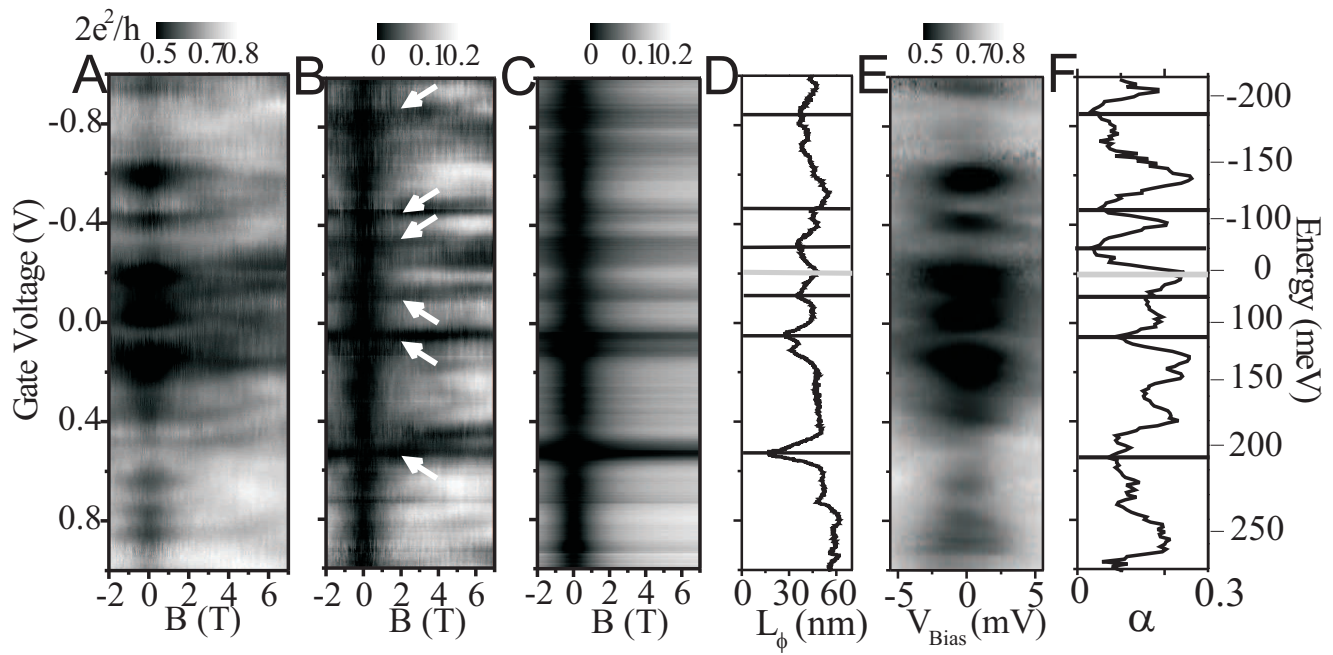}
\caption{\label{fig2}(A) Grey scaled conductance $G$ of sample A as a
  function of gate voltage $U$ and perpendicular magnetic field at a
  temperature of 10 K. (B) Deviation of $G$ from the zero-field conductance $G(U,B)-G(U,0)$. White arrows indicate the regions of quenched magnetoconductance. (C) Reproduction of the magnetoconductance by 1D weak localization fits. The parameters $L_{\varphi}$ and $G(B=0)$ are used as obtained by fitting the data on Fig. 2A. (D) Phase coherence length $L_{\varphi}$ vs. gate voltage as obtained from the fit. The positions of the charge neutrality point (grey line) and the regions of quenched magnetoconductance (black lines) are indicated. (E) Differential conductance of sample A as a function of gate voltage and dc bias voltage $V_{\rm Bias}$ at $T$=10 K. (F) Exponent $\alpha$ vs. gate voltage as obtained from fitting a power law $V^{\alpha}$ to the differential conductance in the range $eV \gg k_{\rm B}T$. Right: scale conversion of the gate voltage into a (nonlinear) energy scale using the gate capacitance as obtained from Fig. 3B.}
\end{figure*}
 We have checked for several gate voltages that $G(B)$ is symmetric
 with respect to magnetic field reversal as required in a two point
 configuration (not shown). In addition, most of the curves show a
 conductance minimum at zero magnetic field. A closer look at the data
 reveals that both the amplitude and the width of the conductance dip
 vary strongly with gate voltage. In order to make this variation more
 visible, we subtracted the curve at zero magnetic field (see Fig. 1C)
 from all gate traces at finite fields. The deviation from the
 zero-field conductance is presented as a color plot in Fig. 2B. The
 most striking observation is that the magnetoconductance (MC)
 disappears at certain gate voltages $U^*$, as indicated by
 arrows. These voltages $U^*$ are grouped symmetrically around the
 conductance minimum at \mbox{$U_{\rm Gate}\approx -0.2$ V} in
 Fig. 1B, which we have assigned to the charge neutrality point. The
 position of the latter, as well as the gate voltages of MC quenches
 have been indicated also in the linear response conductance curve
 (Fig 1C) by red and black vertical lines, respectively. The latter
 always coincide with conductance maxima. These observations lead us
 to the conjecture that the quenched MC may occur at the onset of
 subbands of the outermost nanotube shell, which is believed to carry
 the major part of the current at low temperatures
 \cite{Bachtold_AB}.\\ 
To confirm this idea, we applied a simple bandstructure model. The
 black line in Fig. 3A shows the density of states of a single wall
 (140,140) armchair nanotube, which matches to the diameter of sample
 A (19 nm).
\begin{figure}
\includegraphics{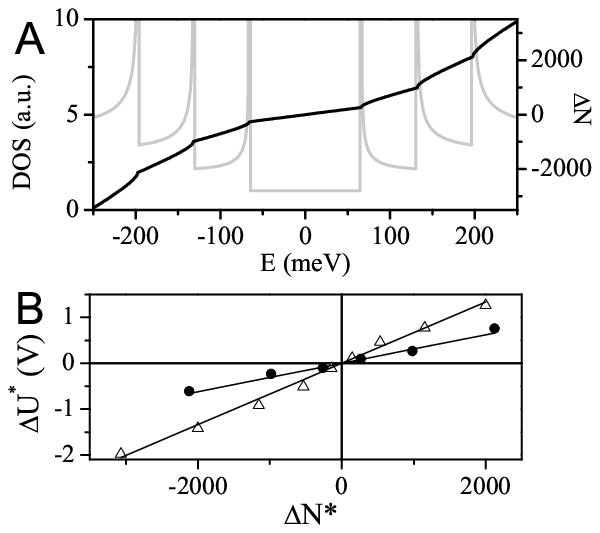}
\caption{\label{fig3}(A) Calculated $\pi$-orbital density of states (DOS) for a (140,140) armchair nanotube of diameter of 19 nm (grey) as a function of energy. Number of excess electrons $N(E)$ (black) as obtained from the integration of the DOS from 0 to $E$. The subband spacing for this diameter is 66 meV. (B) Measured gate voltage values $U^*$ of nanotube subband onsets vs. calculated numbers of electrons $\Delta N^*$ at subband onsets for sample A (circles, diameter 19 nm) and B (triangles, diameter 14 nm). The lines correspond to linear fits of the data. The slopes of the lines correspond to gate capacitances per length of 120 aF/$\mu$m and 129 aF/$\mu$m for sample A and B, respectively.}
\end{figure}
 Typical van Hove singularities arise at the energies, where the
 subband bottoms are located \cite{SaitoDresselhaus}. By integration
 over energy one obtains the number $\Delta N$ of excess electrons on
 the tube, plotted as a red line in Fig. 3A. In this way, we can
 determine the number $\Delta N^*$ of electrons at the onset of the
 nanotube subbands. If we assume as usual a capacitative coupling
 between the gate and the tube, $\Delta N$ can be converted into a
 gate voltage via $CU_{\rm Gate}=e\Delta N$. In Fig. 3B the measured
 gate voltages $U^*$ of quenched MC are plotted versus the calculated
 $\Delta N^*$ for both samples. Both data sets fit very well into
 straight  lines, which demonstrates that most of the positions $U^*$
 of the quenched MC agree very well with the expected subband
 onsets. In addition, the gate capacitances $C$ are provided by the
 slope of $U^*$ vs. $\Delta N^*$. The capacitances per length are
 nearly identical, i.e. 120 aF/$\mu$m and 129 aF/$\mu$m for samples A
 and B, respectively. These values agree within a factor of 2 with
 simple geometrical estimates of $C$, indicating the consistency of
 the interpretation. From the capacitance $C$ and the calculated
 dependence of the number of electrons $N$ on energy one can convert
 the gate voltage into an equivalent Fermi energy. This energy scale
 is shown in Fig. 2F.\\ 
The typical dip in the MC at $B=0$ in Fig. 2A has been observed
earlier and can be explained in terms of weak localization in absence
of spin-orbit scattering \cite{Schoenenberger, LiuAvouris,
  Tarkianinen}. The weak localization correction $\Delta G_{\rm WL}$
to conductance provides information on the phase coherence length
$L_{\varphi}$ of the electrons. With $W$ being the measured diameter
and $L=300$ nm the electrode spacing of the nanotube, $\Delta G_{\rm
  WL}$ is given in the quasi-one-dimensional case \mbox{($L_{\varphi}
  > W$)} by $\Delta G_{\rm WL}=-(e^2/\pi \hbar L)\cdot
(L_{\varphi}^{-2}+W^2/3\ell_m^4)^{-1/2}$, where $\ell_{\rm m}=(\hbar
/eB)^{1/2}$ is the magnetic length. In Fig. 2B each row displays a dip
around zero magnetic field, where both the amplitude and the width of
the dip vary strongly with gate voltage. We have used the weak
localization expression above to fit the low field MC with
$L_{\varphi}$ and $G(B=0)$ as free parameters. The conductance $\Delta
G_{\rm WL}$ as calculated using the fit parameters is plotted in
Fig. 2C. We find that conductance traces are reproduced very well by
the fit for fields up to 2 T. For higher fields, deviations occur,
most probably due to residual universal conductance fluctuations. In
this way we obtain an energy dependent  phase coherence length
$L_{\varphi}(E_{\rm F})$, which is plotted in Fig. 2D.  $L_{\varphi}$
varies from 20 to 60 nm and displays pronounced minima which
correspond to the regions of nearly flat MC in Fig. 2B. From the
preceding discussion, we can say that weak localization seems to be
suppressed at the onset of nanotube subbands.\\ 
In order to confirm the validity of our interpretation in terms of
 weak localization, we have studied the temperature dependence of the
 phase coherence length. As the dominating dephasing mechanism,
 quasielastic electron-electron scattering has been identified
 \cite{Schoenenberger,LiuAvouris,Stojetz}. Dephasing by
 electron-phonon scattering is negligible since the corresponding mean
 free path exceeds \mbox{1 $\mu$m} even at \mbox{300 K}
 \cite{park,javey}. The theory by Altshuler, Aronov and Khmelnitzky
 \cite{Altshuler_Aronov_Khmelnitzky} predicts $L_{\varphi}=(GDL\hbar
 ^2/2e^2k_{\rm B}T)^{1/3}$, where $G$ is the conductance, $D$ is the
 diffusion constant, $L$ is the length of the tube. The dominance of
 electron-electron-scattering can be confirmed by studying the
 temperature dependence of  $L_{\varphi}$. Therefore, the MC
 measurements have been repeated for temperatures ranging from 1 K to
 60 K. In order to eliminate the contribution of the universal
 conductance fluctuations, the MC curves have been averaged over all
 gate voltages. The result is plotted in Fig. 4A.
\begin{figure}
\includegraphics{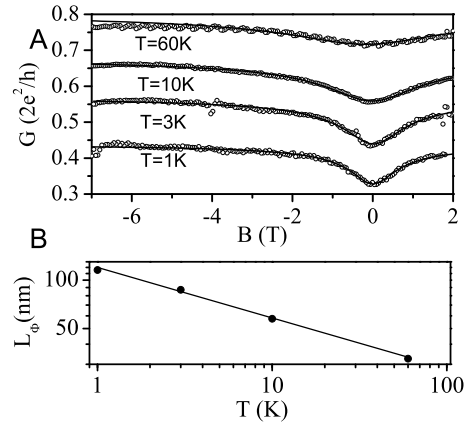}
\caption{\label{fig4} (A) Averaged magnetoconductance of sample A (circles) at temperatures of 60 K, 10 K, 3 K and 1K (top to bottom) and fits of 1D weak localization behavior (lines). (B) Double-logarithmic plot of the temperature dependence of the phase coherence length $L_{\varphi}$ as obtained from the weak localization fit (black dots). The line corresponds to a power law fit with an exponent -0.31.}
\end{figure}
 For the comparison of the curves with theory, one has to bear in mind
 that the average runs also on curves with suppressed MC. Hence, for
 the fit an averaged weak localization contribution of the form
 $\Delta G_{\rm WL}^*=A\cdot \Delta G_{\rm WL}$ with a scaling factor
 $0<A<1$ has been taken into account. The fitted curves are included
 in Fig. 4A. They match the data very well, up to magnetic fields of 7
 T. In Fig. 4B the resulting  $L_{\varphi}(T)$ are presented. The
 contribution of the universal conductance fluctuations is completely
 suppressed by ensemble averaging. The temperature dependence matches
 a power law with exponent -0.31, which is close to the theoretical
 prediction of -1/3.\\  
Another quantum correction to the conductance is induced by the
electron-electron-interaction and reduces the density of states near
the Fermi energy  \cite{EggerGogolin}. This leads to zero bias
anomalies in the differential conductance d$I$/d$V$
\cite{Bachtold_Tunnel}, from which information on the strength of the
electron-electron-interaction can be extracted. In the case of
tunneling into an interacting electron system with an ohmic
environment, the differential conductance d$I$/d$V$ is given by a
power law, i.e. d$I$/d$V\propto V^{\alpha}$ for $eV \gg k_{\rm B}T$,
where the exponent $\alpha$ depends both on the interaction strength
and the sample geometry \cite{FisherKane}. In order to obtain
complementary information, we have examined the dependence of the ZBA
on the gate voltage $U_{\rm Gate}$. The differential conductance has
been measured as a function of $U_{\rm Gate}$ and $V_{\rm Bias}$. The
result is presented in Fig. 2E. For each gate voltage, the conductance
shows a dip at zero bias. The zero bias anomaly has a strongly varying
width with gate voltage and nearly vanishes at the same gate voltages
$U_{\rm Gate}=U^*$ as the magnetoconductance. For each value of the
gate voltage, a power law fit for the bias voltage dependence of the
differential conductance has been performed. The resulting exponent
$\alpha(U_{\rm Gate})$ is plotted in Fig. 2F. $\alpha$ varies between
0.03 and 0.3 and shows pronounced minima at the gate voltages $U^*$.
\\ 
We thus observe experimentally a strong correlation between the single
particle interference effects (expressed by $L_{\varphi}$) and the
interaction effects (expressed by $\alpha$). Both are strongly reduced
at certain positions of the Fermi level, which match well the
positions of the van Hove singularities estimated from simple
bandstructure models. What is the effect of the bandstructure?
Numerical calculations by Triozon et al. \cite{Triozon} indicate that
the diffusion coefficient $D$ is not a constant as a function of
$E_F$, but displays pronounced minima at the onset of new subbands. At
these points strong scattering occurs, resulting from the opening of a
highly efficient scattering channel. This has a direct effect on
$L_{\varphi}=\sqrt{D(E_F)\tau_{\varphi}}$. Of course, $\tau_{\varphi}$
may also be affected.\\ 
Can the energy dependence of $D(E_{\rm F})$ also explain the
suppression of the interaction effects? This question has already been
raised by Kanda et al. \cite{Kanda}, who also observed a pronounced
gate modulation of $\alpha$. For weak electron-electron-interaction
the theory of Ref. \cite{EggerGogolin} predicts $\alpha \propto
1/\ell_{\rm el}$, where $\ell_{\rm el}$ is the elastic mean free
path. This is definitely incompatible with the observed suppression of
$\alpha$ at Fermi levels where diffusion is slow. The observed strong
modulations of $L_{\varphi}$ and $\alpha$ are accompanied by a rather
weak modulation of the zero bias conductance at 10 K (see
Fig. 1B). One may thus ask, whether the assumption of weak
interactions is valid. Taking the simple Drude formula $\sigma =
e^2N(E_{\rm F})D(E_{\rm F})$ as an orientation, this can be explained
by a partial compensation of the variation of $N$ and $D$ with $E_{\rm
  F}$. However, a quantitative explanation of the observed interplay
between bandstructure effects and quantum corrections to the
conductance requires a realistic model calculation for a thick, e.g.,
(140,140) nanotube including disorder and interaction effects. The
simple model of strictly one-dimensional conductance channels is
obviously incompatible with the observed weak-localization-like
magnetoconductance close to the charge neutrality point. The disorder
must be strong enough to mix the channels without completely smearing
the density of states. \\    
In conclusion, our electronic transport measurements on multiwall
carbon nanotubes reveal an interplay of bandstructure effects
originating from the geometry of the tube and quantum interference
induced by disorder. The results demonstrate the necessity of a
systematic theoretical approach which can account both for disorder
and geometrical effects on the same level.
\begin{acknowledgments}
 We have benefitted from inspiring discussions with A.~Bachtold, V.~Bouchiat, G.~Cuniberti, H.~Grabert, M.~Grifoni, K.~Richter, S.~Roche, R.~Sch\"afer, C.~Sch\"onenberger and F.~Triozon. Funding by the Deutsche Forschungsgemeinschaft within the Graduiertenkolleg 638 is acknowledged. The work in Lausanne was supported by the Swiss National Science Foundation.
\end{acknowledgments}
       

\begin{thebibliography}{1}
\bibitem{DekkerReview}
C.~Dekker, {\it Physics Today} {\bf 52} 22 (1999).

\bibitem{Schoenenberger} 
C.~Sch\"onenberger {\it et al.}, {\it Appl. Phys. A} {\bf 69} 283 (1999).

\bibitem{Frank} 
S.~Frank  {\it et al.}, {\it Science} {\bf 280}  1744 (1998).

\bibitem{Urbina}
A.~Urbina {\it et al.}, {\it Phys. Rev. Lett.\/} {\bf 90}, 106603 (2003).


\bibitem{Langer}
L.~Langer {\it et al.}, {\it Phys. Rev. Lett.\/} {\bf 76}, 479 (1996).

\bibitem{LiuAvouris} 
K.~Liu {\it et al.}, {\it Phys. Rev. B} {\bf 63} 161404 (2001).

\bibitem{Bachtold_AB}
A.~Bachtold {\it et al.}, {\it Nature} {\bf 397}, 673 (1999).

\bibitem{Bachtold_Tunnel}
A.~Bachtold {\it et al.}, {\it Phys. Rev. Lett.} {\bf 87}, 166801 (2001).

\bibitem{SaitoDresselhaus}
R.~Saito {\it et al.}, {\it J. Appl. Phys.} {\bf 73} 494 (1993).

\bibitem{TansDekker393} 
S.~J.~Tans  {\it et al.}, {\it Nature} {\bf 393} 49 (1998).

\bibitem{Krueger}
M.~Kr\"uger {\it et al.}, {\it New J. Phys} {\bf 5} 138.1 (2003).

\bibitem{Bockrath}
P.~L.~McEuen  {\it et al.}, {\it Phys. Rev. Lett.\/} {\bf 83}, 5098 (1999).


\bibitem{Tarkianinen}
R.~Tarkianinen {\it et al.}, {\it Phys. Rev. B} {\bf 64}, 195412 (2001).


\bibitem{Stojetz}
B.~Stojetz {\it et al.}, {\it New J. Phys} {\bf 6} 27 (2004).

\bibitem{park}
J.-Y.~Park {\it et al.}, {\it Nano Lett.} {\bf 4}, 517 (2004).

\bibitem{javey}
A.~Javey {\it et al.}, {\it Phys. Rev. Lett.} {\bf 92}, 106804 (2004).



\bibitem{Altshuler_Aronov_Khmelnitzky}
B.~L.~Altshuler, A.~G.~Aronov and D.~Khmelnitzky, {\it Solid State Comm.\/} {\bf 39}, 619 (1981).

\bibitem{EggerGogolin}
R.~Egger and A.~O.~Gogolin, {\it Phys. Rev. Lett.} {\bf 87}, 066401 (2001).

\bibitem{FisherKane}
C.~L.~Kane and M.~P.~W.~Fisher, {\it Phys. Rev. B} {\bf 46}, 15233 (1992).

\bibitem{Triozon}
F.~Triozon {\it et al.}, {\it Phys. Rev. B\/} {\bf 69}, 121410 (2004).

\bibitem{Kanda}
A.~Kanda {\it et al.}, {\it Phys. Rev. Lett.} {\bf 92}, 026801 (2004).
\end{thebibliography}

\end{document}